\documentclass{ws-procs9x6-cpt19}
\begin{document}

\newcommand{\refeq}[1]{(\ref{#1})}
\def\etal {{\it et al.}}

\title{Probing Lorentz Invariance \\
With Top Pair Production at the LHC and Future Colliders}

\author{A.\ Carle, N.\ Chanon, and S.\ Perri\`es}

\address{Universit\'e de Lyon,
Universit\'e Claude Bernard Lyon 1,\\
CNRS-IN2P3,
Institut de Physique Nucl\'eaire de Lyon,\\
Villeurbanne 69622, France}

\begin{abstract}
This article presents prospects for Lorentz-violation searches
with $t\bar{t}$ at the LHC and future colliders.
After a short presentation of the Standard-Model Extension
as a Lorentz-symmetry-breaking effective field theory,
we will focus on $t\bar{t}$ production.
We study the impact of Lorentz violation
as a function of center-of-mass energy
and evaluate the sensitivity of collider experiments to this signal.
\end{abstract}

\bodymatter

\section{Introduction}

The top-quark sector of Standard-Model Extension (SME) is weakly constrained.
Since the LHC is a top factory,
it provides a unique opportunity to search for Lorentz violation (LV).
The SME is an effective field theory
including all LV operators.
Here,
we consider the LV CPT-even part of the lagrangian
modifying the top-quark kinematics:\cite{Kostelecky}
\begin{equation}
\mathcal{L}^\text{SME} \supset \frac{i}{2} (c_L)_{\mu \nu} \bar{Q}_t \gamma^\mu \overset{\leftrightarrow}{D^\nu} Q_t + \frac{i}{2} (c_R)_{\mu \nu} \bar{U}_t \gamma^\mu \overset{\leftrightarrow}{D^\nu} U_t,
\end{equation}
where $Q_t$ and $U_t$ denote the left- and right-handed top-quark spinors,
respectively. 
The $c_{\mu\nu}$ coefficients are constant in an inertial frame,
taken to be the Sun-centered frame.
We aim at measuring the constant coefficients:\cite{KosteleckyPheno}
\begin{equation}
c_{\mu \nu} = \frac{1}{2} \left[
(c_L)_{\mu \nu} + (c_R)_{\mu \nu}
\right],
\qquad d_{\mu \nu} = \frac{1}{2} \left[
(c_L)_{\mu \nu} - (c_R)_{\mu \nu}
\right].
\end{equation}
Expressions for these coefficients
in a laboratory frame on Earth
will introduce a time dependence of the cross section for $t\bar{t}$ production
owing to the Earth's rotation around its axis.
This time dependence can be exploited to search for LV at hadron colliders.

To express the $c_{\mu\nu}$ coefficients
in the reference frame of a hadron circular collider,
we need:
\begin{itemize}
\item the latitude $\lambda$, i.e., the angle between the equator and the poles,
\item the azimuth $\theta$,\cite{Jones} i.e., the angle between the Greenwich tangent vector and the clockwise ring collider tangent vector,
\item the longitude impacts only the phase of the signal because of the Earth's rotation around its axis, and
\item the Earth's angular velocity $\Omega$.
\end{itemize}

\section{Modulation of the $t\bar{t}$ cross section}

The analysis aims at measuring the time dependence of the $t\bar{t}$ cross section
\begin{equation}
\sigma_\text{SME} = \left[1 + f(t)\right]\sigma_\text{SM}.
\end{equation}
A first analysis of this kind
was performed with the D0 detector at the Tevatron.\cite{D0}
We use here the same benchmarks.
We analyze Wilson's coefficients
for a couple of non-null $c_{\mu \nu}$:
$c_{XX} = -c_{YY}$,
$c_{XY} = c_{YX}$,
$c_{XZ} = c_{ZX}$ or $c_{YZ} = c_{ZY}$.
Each of these scenarios generates an oscillating behavior of the amplitude.
The latitude $\lambda$ and the azimuth $\theta$ affect the amplitude
while the Earth's angular velocity $\Omega$ affects the frequency.
In the case of $c_{XX} = -c_{YY}$ and $c_{XY} = c_{YX}$,
$f(t)$ has a period of one sidereal day.
On the other hand,
in the $c_{XZ} = c_{ZX}$ and $c_{YZ} = c_{ZY}$ case,
the amplitude has a period of one half of a sideral day.
More detailed expressions are given in Refs.\ \refcite{KosteleckyPheno,D0}.

\section{Expected sensitivity}

In this work,
samples of $t\bar{t}$ with dilepton decay
were generated with MadGraph-aMC@NLO 2.6.
It was found that the amplitude of the LV $t\bar{t}$ signal
is increasing with the center-of-mass energy.
The signal amplitude
as a function of the center-of-mass energy in $p$--$p$ collisions
(with CMS or ATLAS as the laboratory frame)
increases from $0.001$ at D0
(in the $c_{XY} = c_{YX} = 0.01$ scenario)
to $0.045$ at the LHC Run II ($13\,$TeV)
and to $0.055$ at the Future Circular Collider (FCC, $100\,$TeV).

We evaluate the expected sensitivity to the signal for each benchmark.\cite{Carle}
As a consequence of the increase in luminosity,
the increase in cross section,
and the increase in the amplitude of the LV signal,
we find the following expected sensitivities to the SME coefficient
$c_{\mu\nu}$ in the $c_{XX} = -c_{YY}$ case:
\begin{itemize}
    \item $\Delta c = 7\times 10^{-1}$: D0 ($\sqrt{s} = 1.96\,$TeV,
    $\mathcal{L} = 5.3\,$fb$^{-1}$),
    \item $\Delta c = 1\times 10^{-3}$: LHC Run II ($\sqrt{s}$ = $13\,$TeV, $\mathcal{L} = 150\,$fb$^{-1}$),
    \item $\Delta c = 2\times 10^{-4}$: HL-LHC ($\sqrt{s} = 14\,$TeV,
    $\mathcal{L} = 3000\,$fb$^{-1}$),
    \item $\Delta c = 3\times 10^{-5}$: HE-LHC ($\sqrt{s} = 27\,$TeV,
    $\mathcal{L} = 15\,$ab$^{-1}$),
    \item $\Delta c = 9\times 10^{-6}$: FCC ($\sqrt{s} = 100\,$TeV,
    $\mathcal{L} = 15\,$ab$^{-1}$).
\end{itemize}

\section{Signal amplitude at hadron colliders}

\begin{figure}
\begin{center}
\includegraphics[width=3.0in]{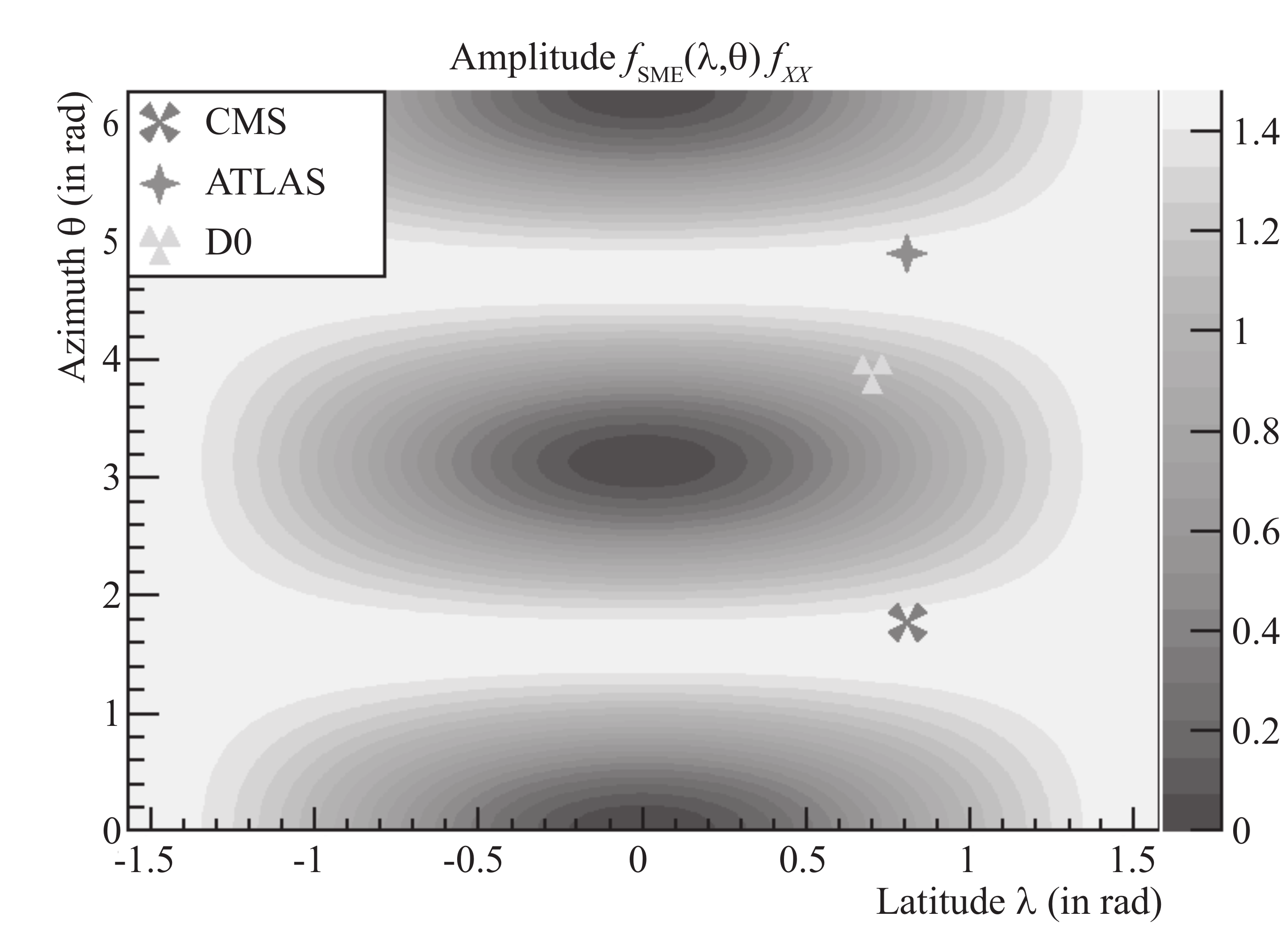}
\end{center}
\caption{Amplitude of $f(\lambda, \theta)$ as a function
of latitude and azimuth
for the $XX$, $YY$, and $XY$ benchmarks.}
\label{aba:fig4}
\end{figure}

A noticeable fact is the dependence of the signal amplitude
on the latitude and azimuth of the collider experiment on Earth.
This dependence is presented in Fig.\ \ref{aba:fig4}.
We find that performing such an experiment at the LHC
would increase the sensitivity to SME coefficients in the top sector
by two orders of magnitude.
Further improvements are expected at future colliders.

\section*{Acknowledgments}
Thanks to Alan Kosteleck\'y and Ralf Lehnert
for giving me the opportunity to expose my work
in front of a benevolent and hard-working community.

\end{document}